\documentclass[11pt]{article}
\usepackage{epsfig}
\begin{document}   
\begin{center}
{ \small Reports of the National Academy of Sciences of Ukraine 2000, 10, p.36-40 }
\end{center}
\begin{center}{ \bf  Opportunity of representation of the nonlinear wave equations
through a variable action-angle   }
\end{center}
\begin{center}{  A.N. Skripka }
\end{center}
\begin{center}{  KCK Soft, Kiev, Ukraine }
\end{center}
\begin{center}{ \small skripka@ukrpost.net }
\end{center}

{ \it The approach allowing is considered to represent
the solutions such as stationary lonely waves of various nonlinear 
wave the equations as system of the ordinary differential equations 
in variable action - angle.  
}

\vspace*{3mm}

1. Introduction

In the nonlinear theory of oscillations the essentially important meaning 
has an opportunity transition in researched system from initial variable to variable 
action - angle. If oscillatory system with two degrees of freedom 
is integrable hamiltonian system,
it is possible transformed to a kind  
$$
     \frac{dI}{dt}=0,\qquad  \frac{d\Theta}{dt}=\omega_0(I),\qquad
     I=\frac{1}{2\pi}\oint pdq\eqno(1), 
$$
where $ I=const $ - action, $ (q(t),p(t)) $ - canonical variable 
of hamiltonian system, $ \omega_0 (I) $ - frequency of oscillations.

If to consider only oscillatory systems with two degree of freedom
(on a phase plane), in a case hamiltonian systems, the action is 
the area limited on a phase plane to a trajectory of system [1]. 
Action and frequency of oscillations are determined, how 
in parameters of system, and initial conditions. If to consider
nonhamiltonian (dissipative)  system, stationary periodic
( on an infinite interval of time)  oscillations in them, correspond one or
more limiting cycles, and, accordingly, as much of pairs of meanings variable
action - angle. In a case, when a limiting cycle on plane, the action is equal to the area
limited on a phase plane by a limiting cycle [2].

From nonlinear oscillations this formalism can be distributed to some
kinds of nonlinear waves. In work [3] is shown, that basic
nonlinear wave models having solutions as solitons (Korteveg-de Vriz equation, 
sin-Gordon equation and some other) are complete integrable systems and them it is 
possible to transform to the equations to a variable action - angle.

Opportunity of representation in a variable action - corner will be considered below
the solutions such as stationary lonely waves for some kinds nonlinear 
the wave equations, that is actually in an implicit kind is carried out 
transformation from difficult nonlinear partial differential equation 
to the elementary system the ordinary differential equations.

2. The Korteveg-de Vriz equation
$$
     \frac{\partial u}{\partial t}+Au\frac{\partial u}{\partial x}+
     \frac{\partial^3 u}{\partial x^3}=0.\eqno(2)
$$
Through substitution such as a running wave
$ u(x,t)=u(z)=u(x-vt), $\linebreak
$ v=const $
- speed of a wave, it is possible to transform  to the ordinary  
differential equation
$$ 
\frac{d^2u}{dz^2}-vu+0.5u^2=0,\eqno(3) 
$$
which is hamiltonian system with canonical variable
$ (u,du/dz) $ and Hamilton function
$$  
H(u,du/dz)=0.5((du/dz)^2-vu^2+Au^3/3). 
$$
To the solution as a lonely wave of the Korteveg-de Vriz equation (2) corresponds
the solution as the closed loop of separatrix ( fig. 1 ) equation (3): 
$$ 
u(z)=3vA^{-1}sech^2(0.5\sqrt vz).\eqno(4) 
$$
Proceeding from equality (1), action on each closed trajectory 
equation (3) is equal:
$$ I=\frac{1}{2\pi}\oint pdq=\frac{1}{2\pi}\oint (du/dz)du=
  \frac{1}{2\pi}\oint(du/dz)^2dz=const.
$$
Separatrix also is the closed trajectory, as
$ u(z\to\pm\infty)\to 0. $
On separatrix
$$ I=\frac{1}{2\pi}\int_{-\infty}^{\infty}(du/dz)^2dz=I_1=const,\qquad
   \frac{dI}{dz}=0,
$$
where  
$ u(z) $ 
is defined by equality(4).

Let's analyse sense a variable angle for the description stationary lonely waves. 
In case of stationary periodic oscillations variable angle describes phase of oscillations, 
that is complete angle on which the system an initial situation
has deviated for an interval 
time from the moment of a beginning of oscillations. By analogy it is possible to assume, 
that in case of stationary lonely waves the variable angle will be to describe distance, 
on which the wave from has left initial situation
and complete derivative of a angle is speed of a wave:
        $$ \frac{d\Theta}{dz}=v $$   

Thus in case of dynamics such as a lonely stationary wave the Korteveg - de Vriz 
equation can be transformed to system of the equations in a variable action - angle:
$$  \frac{dI}{dz}=0,\qquad  \frac{d\Theta}{dz}=v,\qquad
    I=\frac{1}{2\pi}\int_{-\infty}^{\infty}(du/dz)^2dz,
$$
where 
$ z=x-vt,\qquad u(z\to\pm\infty)\to 0. $ 
The speed $ v $ is any real number also is determined by the initial conditions 
and parameters of the equation (2).
\\

3. The sin-Gordon equation
$$
  \frac{\partial^2 u}{\partial t^2}+\sin u=\frac{\partial^2 u}
       {\partial x^2}.\eqno(5)
$$
Through substitution such as a running wave 
$ u(x,t) =u(z) =u(x-vt), v=const $ 
it can be transformed to the ordinary differential equation
$$ \frac{d^2u}{dz^2}+(v^2-1)^{-1}\sin u=0,\eqno(6) $$ 
which is hamiltonian system with canonical variable 
$ (u, du/dt) $. 
To the solution as lonely waves of the sin-Gordon equation corresponds 
solution as top or bottom 
halfloop of separatrix (fig. 2) of the equations (6): 
$$ 
(top\quad halfloop): u(z)=4arctg\exp[z(1-v^2)^{-1/2}],
$$
$$
 u(z\to -\infty)\to 0,u(z\to +\infty)\to 2\pi.
$$
Value of action is equal
$$
I=\frac{1}{2\pi}\int_{-\infty}^{\infty}(du/dz)^2dz=I_2=const,
$$
and the equation (5) will be transformed to a kind
$$  \frac{dI}{dz}=0,\qquad  \frac{d\Theta}{dz}=v, $$
where $ u(z\to -\infty)\to 0,u(z\to +\infty)\to 2\pi,z=x-vt.$
The speed $ v $ is determined by the initial conditions and parameters 
of equation(5).
\\

4. The Kolmogorov-Petrovsky-Piskunov equation.
$$
\frac{\partial u}{\partial t}=D\frac{\partial^2 u}{\partial x^2}+ku(1-u).
\eqno(7)
$$
By substitution such as a running wave
$ u(x,t)=u(z)=u(x-vt),v=const $
it can be transformed to the ordinary differential equation
$$ 
    D\frac{d^2u}{dz^2}+v\frac{du}{dz}+ku(1-u)=0,\eqno(8) 
$$
To the solution such as a running wave of the equation (7) there 
corresponds the solution of a type halfloop of separatrix (fig. 3) 
equation(8) [4]: 
$$ 
   u(z)=(1+\exp(az))^{-2},a=\pm\sqrt{k/(6D)},v=5aD
$$
with the boundary conditions    
$ u(z\to -\infty)\to 0,u(z\to +\infty)\to 1 $.
Value of action is equal
$$
I=\frac{1}{2\pi}\int_{-\infty}^{\infty}(du/dz)^2dz=I_3=const,
$$
and equation (7) will be transformed to a kind
$$
\frac{dI}{dz}=0,\qquad  \frac{d\Theta}{dz}=v=\pm\sqrt{\frac{25kD}6}.
$$
\\

5. The Burgers equatian
$$
     \frac{\partial u}{\partial t}+u\frac{\partial u}{\partial x}=
     D\frac{\partial^2 u}{\partial x^2}.\eqno(9)
$$
By substitution such as a running wave
 $ u(x,t)=u(z)=u(x-vt),v=const $
it can be transformed to the ordinary differential equation,
which under boundary conditions 
$ u(z\to\infty)\to u_1,u(z\to -\infty)\to u_2,u_1<u<u_2 $ 
has the solution [2]:
$$ 
  u(x,t)=u_1+\frac{u_2-u_1}{1+\exp(0.5(u_2-u_1)(x-vt)/D)},v=\frac{u_1+u_2}2.
$$
Value of action is equal
$$
I=\frac{1}{2\pi}\int_{-\infty}^{\infty}(du/dz)^2dz=I_4=const,
$$
and equation (9) will be transformed to a kind
$$
\frac{dI}{dz}=0,\qquad  \frac{d\Theta}{dz}=v=\frac{u_1+u_2}2.
$$

\vspace*{2mm}

\noindent
{\footnotesize
1. Landau L.D.,Lifshitz E.M. Mechanics.-- Moscow: Nauka, 1965. -- 204p.
(in Russian)\\
2. Zaslavsky G.M., Sagdeev R.Z. Introduction in nonlinear physics.--
Moscow: Nauka, 1988. --  368p.(in Russian) \\
3. Takhtadzhian L.A., Faddeev L.D. The hamilton approach in the soliton 
theory.--Moscow: Nauka, 1986. --  528p.(in Russian) \\
4. Gudkov V.V. The solutions such as a lonely wave for two-component 
reaction-diffusion systems // Journal of calculus mathematics and 
mathematical physics -- 1995. -- {\bf 35}, 4. -- p.615 -- 623. (in Russian)\\
}

\epsfig{file=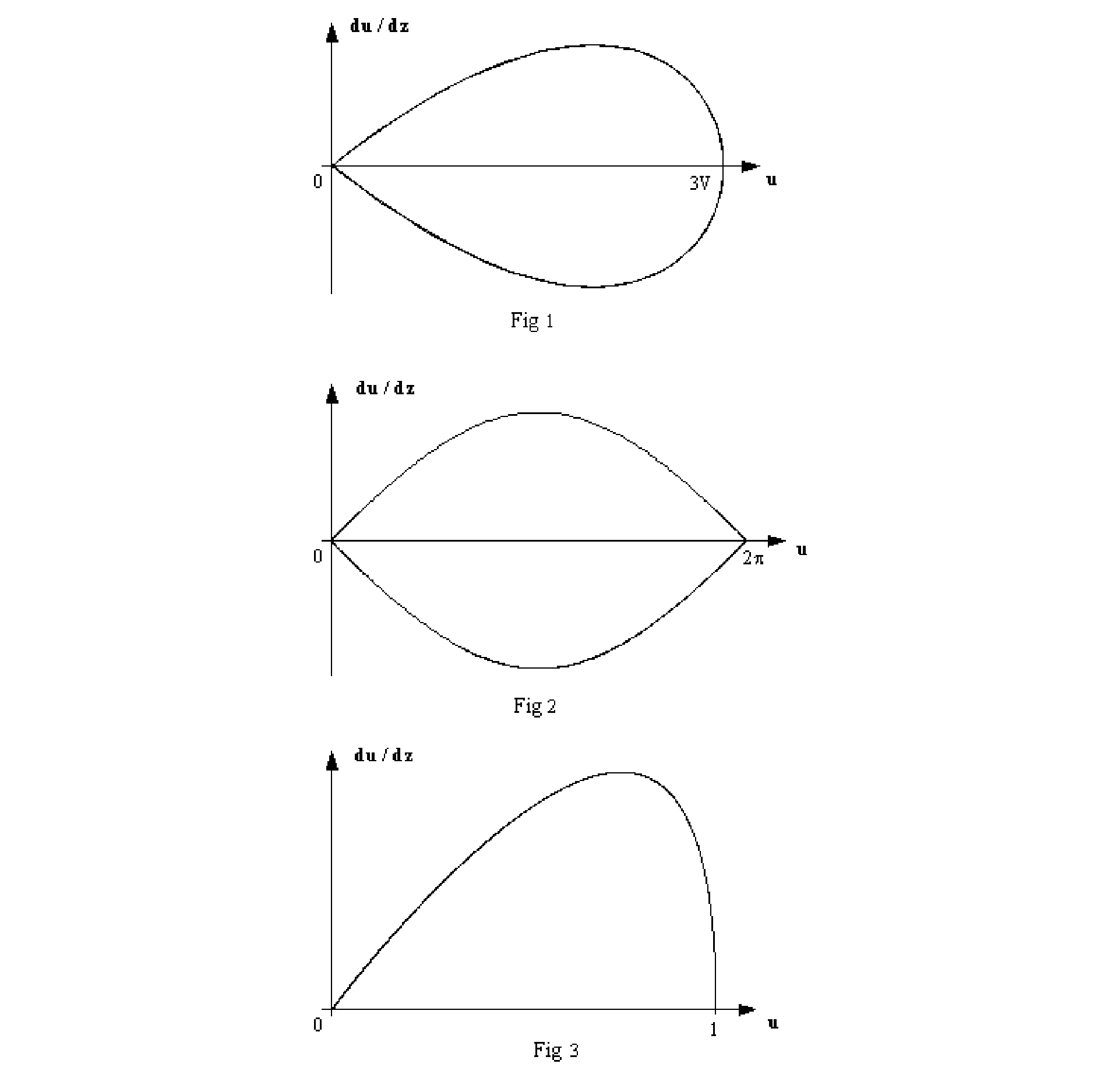, width=\textwidth,height=18cm.}

\end{document}